\documentclass[epj,twocolumn]{webofc}
\usepackage[varg]{txfonts}   
\usepackage{graphicx}

\newcommand{\Teff}{T_{\rm eff}}

\wocname{epj}
\woctitle{Seismology of the Sun and the Distant Stars 2016}
\begin{document}
\title{Probing the Deep End of the Milky Way with New Oscillating {\it Kepler} Giants}
%

\author{\firstname{Savita} \lastname{Mathur}\inst{1}\fnsep\thanks{\email{smathur@spacescience.org}} \and
        \firstname{Rafael A.} \lastname{Garc\'ia}\inst{2} \and
        \firstname{Daniel} \lastname{Huber}\inst{3,4,5} \and
        \firstname{Clara} \lastname{Regulo}\inst{6,7} \and
        \firstname{Dennis} \lastname{Stello}\inst{3,5} \and
        \firstname{Paul G.} \lastname{Beck}\inst{2} \and
        \firstname{Kenza} \lastname{Houmani}\inst{2} \and
        \firstname{David} \lastname{Salabert}\inst{2}
}

\institute{Center for Extrasolar Planetary Systems, Space Science Institute, 4750 Walnut street Suite\#205, Boulder, CO 80301, USA
\and
           Laboratoire AIM, CEA/DRF-CNRS-Universit\'e Paris Diderot; IRFU/SAp, Centre de Saclay, 91191 Gif-sur-Yvette Cedex, France\and
          Sydney Institute for Astronomy (SIfA), School of Physics, University of Sydney, NSW 2006, Australia
\and
	SETI Institute, 189 Bernardo Avenue, Mountain View, CA 94043, USA
\and
	Stellar Astrophysics Centre, Department of Physics and Astronomy, 
Aarhus University, Ny Munkegade 120, DK-8000 Aarhus C, Denmark
\and
	Universidad de La Laguna, Dpto de Astrof\'isica, 38206, Tenerife, Spain
\and
	Instituto de Astrof\'\i sica de Canarias, 38205, La Laguna, Tenerife, Spain
          }

\abstract{%
 The {\it Kepler} mission has been a success in both exoplanet search and stellar physics studies. Red giants have actually been quite a highlight in the {\it Kepler} scene. The {\it Kepler} long and almost continuous four-year observations allowed us to detect oscillations in more than 15,000 red giants targeted by the mission. However by looking at the power spectra of ~45,000 stars classified as dwarfs according to the Q1-16 {\it Kepler} star properties catalog, we detected red-giant like oscillations in ~850 stars. Even though this is a small addition to the known red-giant sample, these misclassified stars represent a goldmine for galactic archeology studies. Indeed they happen to be fainter (down to Kp$\sim$16) and more distant (d>10kpc) than the known red giants, opening the possibility to probe unknown regions of our Galaxy. The faintness of these red giants with detected oscillations is very promising for detecting acoustic modes in red giants observed with K2 and TESS. In this talk, I will present this new sample of red giants with their revised stellar parameters derived from asteroseismology. Then I will discuss about the distribution of their masses, distances, and evolutionary states compared to the previously known sample of red giants.
}
\maketitle
%
\section{Introduction}
\label{intro}

The {\it Kepler} mission \cite{2010Sci...327..977B} has been observing more than 197,000 stars for almost 4 years \cite{2016arXiv160904128M}. While the mission has been a success for its main goal in terms of looking for extrasolar planets and characterizing them (for habitability, occurrence rates etc.) \cite[e.g.][]{2012ApJ...745..120B,2015ApJ...809....8B,2016arXiv160800620K} the excellent quality of the photometric data has enabled us to make tremendous progress on the understanding of stellar evolution and dynamics \cite{2011Natur.471..608B,2012Natur.481...55B,2014A&A...572A..34G,2014A&A...563A..84G,2014A&A...562A.124M,2015Sci...350..423F,2016MNRAS.456..119C,2016Natur.529..181V}. In particular, the mission showed the power of asteroseismology to better constrain the fundamental properties of planet host stars (such as radius and age) needed to characterize planetary systems but also to probe the structure and rotation in deeper layers of stars. 

Some additional studies were also done for binary stars whether to understand their interaction \cite[e.g.][]{2014A&A...564A..36B,2016arXiv160908135P} or to test scaling relations used in asteroseismology \cite[e.g.][]{2016arXiv160906645G}.

More recently, photometric data from CoRoT \cite[Convection, Rotation, and Transits][]{2006cosp...36.3749B} and {\it Kepler} have been combined to large spectroscopic surveys such as APOGEE \cite[Apache Point Observatory Galactic Evolution Experiment][]{2016AN....337..863M} leading to interesting discoveries on chemical abundances evolution of the Galaxy \cite{2015MNRAS.451.2230M,2016arXiv160804951A}, distances of stars \cite{2014MNRAS.445.2758R}.

Here we present the discovery of additional red giants in the {\it Kepler} field that were classified as cool dwarfs in the latest {\it Kepler} star properties catalog of \cite{2014ApJS..211....2H}. For more details this work was presented by \cite{2016ApJ...827...50M} .


\section{Looking at dwarfs}
\label{sec-1}

\begin{figure*}[ht]
\centering
\includegraphics[width=10cm,clip]{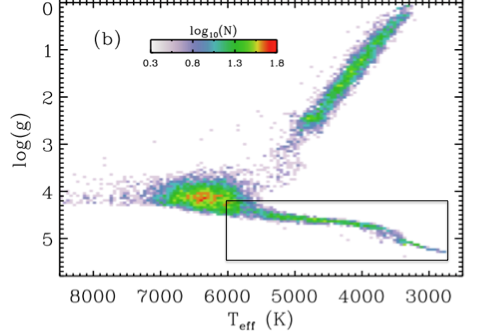}
\caption{Hertzprung-Russell Diagram from the {\it Kepler} star properties catalog of \cite{2014ApJS..211....2H} showing the number density of stars. The box represents the dwarfs selected based on the criteria given in the text for which we visually inspected the power spectra.}
\label{fig-1}       
\end{figure*}

We selected the sample of cool dwarfs based on {\it Kepler} star properties catalog available then \cite{2014ApJS..211....2H} with the following criteria on their gravities and temperatures: $\log g > 4.2$ and $T_{\rm eff} < 6125K$ as shown in Figure~\ref{fig-1} by the box overlaid on the Hertzprung-Russell Diagram from that catalog. This sample was constituted of approximately 45,000 dwarfs. The first purpose of this selection was to study the photon noise of the {\it Kepler} mission as done by \cite{2010ApJ...713L.120J} but using the 17 quarters available to see the temporal variation of the instrumental noise. 

\begin{figure}[htb]
\centering
\includegraphics[width=\hsize,clip]{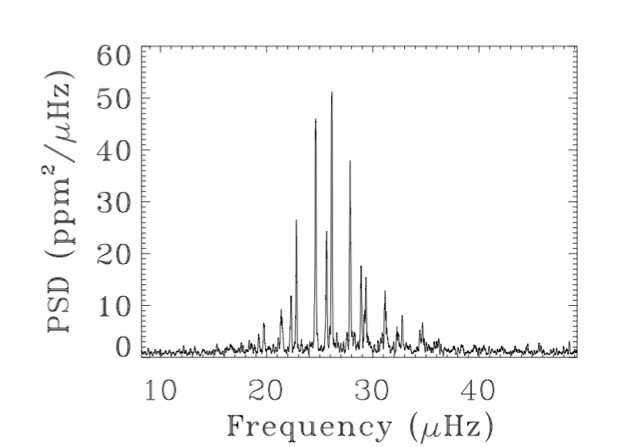}
\caption{Example of the power spectrum density (PSD) of a misclassified red giant, KIC~3216802, showing the excess power of the acoustic modes.}
\label{fig-2}       
\end{figure}

Figure~\ref{fig-2} shows an example of the power spectrum  of one of the stars (KIC 3216802) of this sample that had the following parameters according to the \cite{2014ApJS..211....2H} catalog: 

\begin{itemize}
\item {\it Kepler} magnitude, $Kp$=15.9
\item $\Teff$ = 4745\,$\pm$\,126\,K
\item $\log g$ = 4.8 dex
\item M$\sim$0.5\,M$_\odot$
\item R$\sim$0.48\,R$_\odot$
\end{itemize}

The power spectrum clearly shows an excess power around 25\,$\mu$Hz, which is the region of acoustic modes corresponding to a red giant. The global seismic parameters of this power spectrum suggest that this star has a surface gravity of $\sim$2.38 dex, a mass of 0.73\,$\pm$\,0.3M$_{\odot}$ and a radius of 9.1R$_{\odot}$.  Given this discovery we visually checked the power spectra of the full sample of stars we had selected. We also added some hotter stars that had been observed in short cadence during the survey phase. This led to more than 1000 candidate dwarfs that could be red giants.

\section{Characterizing the misclassified red giants}
\label{sec-3}

We first calibrated the light curves following the KADACS pipeline as described in \cite{2011MNRAS.414L...6G} and then filled gaps shorter than 20 days with the inpainting technique \cite{2014A&A...568A..10G,2015A&A...574A..18P}. We then performed an asteroseismic analysis of the light curves and power spectrum using two different pipelines, A2Z \cite{2010A&A...511A..46M} and SYD \cite{2009CoAst.160...74H} in order to measure the two global acoustic mode parameters: the mean large frequency separation, $\Delta \nu$ and the frequency of maximum power, $\nu_{\rm max}$. 

For the A2Z pipeline, we actually used an improved version called A2Z+ that leads a more precise value of $\Delta \nu$ by applying the auto-correlation and masking the $\ell$=1 modes. Moreover, for stars with modes above 200\,$\mu$\,Hz, they could result from modes that are the reflection of modes above the Nyquist frequency \cite{2014MNRAS.445..946C} so we did a specific analysis to check whether it was the case or not. 

The results from the different pipelines agreed within 1$\sigma$. We re-derived the effective temperatures and distances of the misclassified red giants with an isochrone fitting code using broadband photometry, the asteroseismic observables, and a grid of Parsec isochrones \cite{2012MNRAS.427..127B}.

\section{Checking for pollution}
\label{sec-2}

We then checked for possible pollution that could lead to the detection of acoustic modes in the red-giant regime.
First we looked at the amplitude of the modes. We know there is a tight relation between the maximum amplitude of the modes, $A_{\rm max}$ and the frequency of maximum power, $\nu_{\rm max}$, as showed in \cite{2011ApJ...743..143H}. Stars with mode amplitudes that are lower than the scaling relation derived from a known sample of red giants observed by {\it Kepler} could result from a blending effect diluting the light of the main target. We then checked the J-band images of all the stars and found 36 stars out of the 50 stars with low $A_{\rm max}$ could have a nearby bright star that could pollute the light curves. 

\begin{figure}[h]
\centering
\includegraphics[width=\hsize,clip]{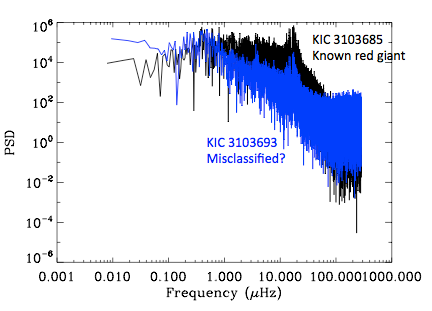}
\caption{Illustration of pollution of a red giant candidate. KIC~3103685 power spectrum (black) from the oscillations of a known red giant, KIC~3103693 (blue), located within 1 arcmin of the main target.}
\label{fig-3}       
\end{figure}

There could also be a nearby red giant that was previously known that could pollute a nearby target. To check for such a phenomena, we compared the power spectra of our red-giant candidates with the power spectra of known red giants located within a radius of 1 arcmin. Figure~\ref{fig-3} shows the superposition of the a red-giant candidate KIC 3103693, superimposed on the power spectrum of a known red giant KIC3103685. We clearly see that in addition of having the same $\nu_{\rm max}$ the shape of the p-mode bell is exactly the same. This means that the misclassified candidate is actually coming from pollution so we discarded all the stars where the power spectra are as similar as in Figure~\ref{fig-3}. 

Finally, we also looked at the crowding, which is the measurement of the target flux compared to the total flux in the optimal aperture as defined by the NASA data. In this definition, a crowding of 1 means that the flux belongs to the main target while a crowding of 0 corresponds to a completely polluted target. Stars with a crowding smaller than 75\% were flagged in the catalog of misclassified stars.

After these checks for possible blends or pollution, we end up with a sample of misclassified red giants of 854 stars.

\section{Conclusions}
\label{sec-con}

\begin{figure*}
\centering
\includegraphics[width=10cm,clip]{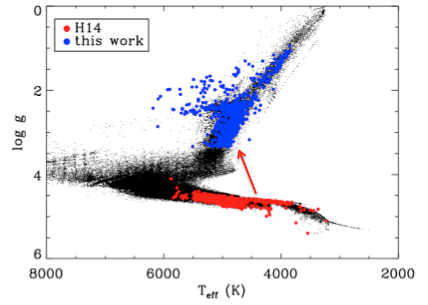}
\caption{HR Diagram of the full sample of {\it Kepler} targets where we superimposed the stars studied in this work with their original parameters (red circles) and with their updated parameters (blue circles). Adapted from \cite{2016ApJ...827...50M}.}
\label{fig-4}       
\end{figure*}

We estimated the surface gravities, masses and radii of these stars with scaling relations based on the Sun \cite[e.g.][]{1995A&A...293...87K,2011A&A...529L...8K} using the new values of effective temperatures. Figure~\ref{fig-4} shows the HR Diagram of the misclassified red giants with their original parameters (blue circles) and the revised ones (red ones). 

We also measured the period spacings of the mixed modes \cite{2011Sci...332..205B,2011A&A...532A..86M} for a subsample of 280 stars allowing us to determine whether the stars are on the Red Giant Branch or Red Clump stars.

The misclassified red giants are in general fainter than the previously known sample from the {\it Kepler} mission, peaking at a magnitude 16 instead o 14. This  proves that asteroseismology can be applied to such faint stars. which is promising for missions such as K2 and TESS.

We also found that misclassified stars are less massive and more distant than the known sample of red giants with distances larger than 10kpc. This sample of stars will thus allow us to probe in more details more distant regions of the Galaxy. 

\begin{figure}[h]
\centering
\includegraphics[angle=90,width=\hsize,clip]{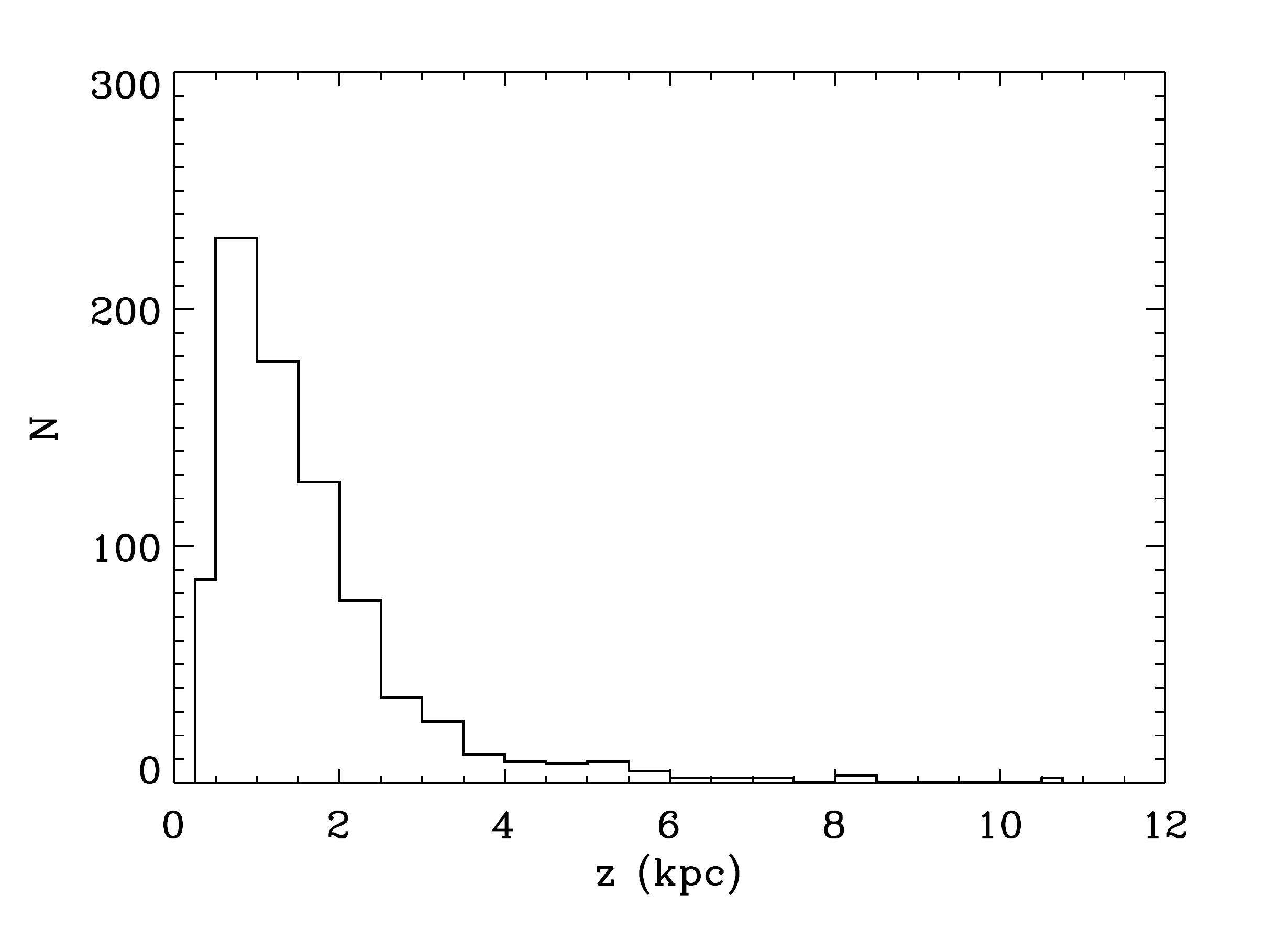}
\caption{Distribution of the heights of the 854 misclassified red giants (from \cite{2016ApJ...827...50M}].}
\label{fig-5}       
\end{figure}

Figure~\ref{fig-5} shows the distribution of the height of the stars compared to the galactic plane. The comparison of our sample with a a distribution from  a synthetic population of halo stars from the Galaxia model \cite{2011ApJ...730....3S} suggests that around 45 red giants of our sample could be halo stars. 

A sample of around 30 stars have already been observed by the APOGEE survey and the spectroscopic analysis agrees with the new seismic classification provided here. The full sample of stars will be studied more thoroughly with the new spectroscopic data for galacto-archeology investigations.

\bibliographystyle{woc} 
\bibliography{/Users/Savita/Documents/BIBLIO_sav} 

%
%
%
%

\end{document}